\documentclass[a4paper]{article}
\usepackage[comma,numbers,square,sort&compress]{natbib}
\renewcommand{\top}{{\mathrm{T}}}
\usepackage{graphicx,amsmath,amssymb}
\usepackage{amsthm,bm}
\theoremstyle{definition} 
\usepackage{bm} 
\newcommand{\lm}{\lambda}
\newcommand{\e}{\varepsilon}
\newcommand{\R}{\mathbb{R}}

\newtheorem{thm}{Theorem}

\newtheorem{lem}{Lemma}
\newtheorem{ass}{Assumption}

\newtheorem{rem}{Remark}

\newcommand{\pb}{\noindent\textbf{Proof. } }
\newcommand{\pe}{\hfill\rule{4pt}{8pt}}

\usepackage{pgf,tikz}
\usetikzlibrary{arrows,shapes,automata,positioning,fit}
\usepackage{pst-node}

\begin{document}
	
\title{Achieving Optimal Output Consensus for Discrete-time Linear Multi-agent Systems with Disturbance Rejection}
	
\author{Yutao Tang, Hao Zhu, and Xiaoyong Lv
	\footnote{
		Yutao Tang is with the School of Artificial Intelligence, Beijing University of Posts and Telecommunications, Beijing. P.\,R. China. Email: yttang@bupt.edu.cn.  Hao Zhu was with the School of Artificial Intelligence, Beijing University of Posts and Telecommunications, Beijing, P.\,R. China, and now is with DiDi Chuxing Technology Company, Beijing, P.\,R. China. Xiaoyong Lv is with the College of Control Engineering, Northeastern University at Qinhuangdao, Qinhuangdao, P.\,R. China. Email: lengfeng0506@163.com.
	}
}

\date{}
	
\maketitle
	
{\noindent\bf Abstract}: In this paper, an optimal output consensus problem is studied for discrete-time linear multi-agent systems subject to external disturbances. Each agent is assigned with a local cost function which is known only to itself. Distributed protocols are to be designed to guarantee an output consensus for these high-order agents and meanwhile minimize the aggregate cost as the sum of these local costs. To overcome the difficulties brought by high-order dynamics and external disturbances, we develop an embedded design and constructively present a distributed rule to solve this problem. The proposed control includes three terms: an optimal signal generator under a directed information graph, an observer-based compensator to reject these disturbances, and a reference tracking controller for these linear agents. It is shown to solve the formulated problem with some mild assumptions. A numerical example is also provided to illustrate the effectiveness of our proposed distributed control laws.

\maketitle

\section{Introduction}
In recent years, a lot of efforts have been made to study the distributed coordination of multi-agent systems. As one of the most important problems, distributed optimization has drawn growing attention due to its wide applications in machine learning, power systems and sensor networks \cite{boyd2011distributed, gharesifard2014distributed, simonetto2014distributed}.  In a typical setting of this problem, a network of interconnected nodes are associated with a group of convex functions, while each node only knows one component of these functions. The design goal is to drive all nodes to reach some steady-states specified by minimizing the sum of these functions through information exchanges with each other. 

Among plenty of publications on this topic, optimal consensus, where the agents are required to reach a consensus on the minimizer of the sum of the local cost functions, has been intensively investigated along with many significant results.   For instance, the authors in \cite{nedic2009distributed} investigated the distributed consensus optimization problem through a novel combination of average consensus algorithms with subgradient methods. Extensions with global or local constraints on the decision variables were further studied in \cite{zhu2012distributed,liu2017constrained}. Efforts have also been made to derive distributed algorithms with fast convergence rate in \cite{jakovetic2014linear,shi2015extra,scaman2018optimal}. Paralleled with these discrete-time results, continuous-time solvers to reach an optimal consensus were also developed under various conditions in \cite{lu2012zero,kia2015distributed,zeng2017distributed}.

At the same time, it is observed that most of the above results are derived only for single-integrator agents from the viewpoint of mathematical programming. In practical applications, the decision variables might be determined by or depend upon physical plants, which can not be described well by single integrators, e.g., a group of mobile robots to achieve a rendezvous \cite{ren2007distributed}. In \cite{tang2018iet}, a numerical example was provided to show that direct use of distributed rules for single integrators might fail to achieve the optimization goal for agents with unity relative degree. Therefore, we should take the high-order dynamics into account  when seeking an optimal consensus in a distributed manner. As the gradient-based rules are basically nonlinear, achieving optimal (output) consensus might be challenging due to the coupling between the high-order dynamics of agents and the distributed optimization requirement. 

Some interesting attempts have been made along this technical line for several kinds of continuous-time high-order dynamics.  For example, the authors in \cite{zhang2017distributed,qiu2019distributed} extended existing distributed optimization rules to continuous-time seconder-order agents by adding some integral terms. Similar ideas have been used in \cite{xie2019global} to achieve optimal consensus for high-order integrators by bounded controls. For a multi-agent system with general linear dynamics, the authors in  \cite{tang2019optimal} proposed an embedded control scheme to solve this kind of optimal coordination problems in a modular way. Some special classes of nonlinear multi-agent systems were also investigated in literature to achieve such an optimal consensus goal in \cite{wang2016distributed,tang2018ijrnc}.  However, in contrast with these papers for continuous-time high-order agents, there is still no general result to our best knowledge on achieving optimal consensus for discrete-time multi-agent systems with non-integrator dynamics. 
	
The objective of this paper is to develop distributed rules for discrete-time high-order agents to achieve an optimal output consensus. To be specific, we assume that the agents are of general linear time-invariant dynamics and can exchange information through a communication topology represented by a directed graph. All agents are to be designed to reach an output consensus and meanwhile minimize the aggregate cost as the sum of local ones. Moreover, we further consider the cases when agents are subject to external disturbances, which are inevitably encountered in practical circumstances.

The contribution of this work is at least two-fold.  On the one hand,  an optimal consensus problem for a group of discrete-time linear multi-agent systems is formulated and solved as a high-order extension of existing discrete-time distributed optimization results for single integrators \cite{lei2016primal,liu2017constrained}.   On the other hand, novel distributed controllers are developed to achieve the optimal (output) consensus goal for these agents under weight-balanced directed graphs with disturbance rejection, which can be taken a discrete-time counterpart of the embedding designs  in \cite{tang2018distributed,tang2019optimal} to solve such optimal consensus problem.  Moreover, the proposed algorithm is free of initialization in contrast with some similar works requiring a nontrivial initialization under both undirected and directed graphs \cite{wang2016distributed, kia2015distributed, xie2019global}, which might be more favorable in large scale multi-agent systems.

The rest of this paper is organized as follows. We first give some preliminaries about graph notations and convex analysis in Section \ref{sec:pre} and then formulate the problem in Section \ref{sec:formulation}. The main design with proofs is presented in Section \ref{sec:main} with a numerical example in Section \ref{sec:simu}. Finally, some concluding remarks are given in Section \ref{sec:con}.

\section{Preliminaries}\label{sec:pre}

In this section, we first provide some preliminaries about graph theory \cite{godsil2001algebraic} and convex analysis \cite{rockafellar1997convex}.

\subsection{Graph theory}

We will use standard notations. Let $\R^N$ be the $N$-dimensional Euclidean space.  $\mbox{col}(a_1,\, \dots,\,a_N) =[a_1^\top,\, \dots,\,a_N^\top]^\top$ for column vectors $a_i\; (i=1,\, \dots,\,N)$. For a vector $a$ (or a matrix $A$), $|a|$ (or $|A|$) denotes its Euclidean (or spectral) norm.  ${\bm 1}_N$ (or ${\bm 0}_N$) denotes an $N$-dimensional all-one (or all-zero) column vector, and $I_{N}$ denotes the $N$-dimensional identity matrix. 

A weighted directed graph (digraph) is described by a triplet $\mathcal {G}=(\mathcal {N}, \mathcal {E}, \mathcal{A})$ with the node set $\mathcal{N}=\{1,{\dots},N\}$ and the edge set $\mathcal {E}$. $(i,\,j)\in \mathcal{E}$ denotes an edge from nodes $i$ to $j$. The weighted adjacency matrix $\mathcal{A}=[a_{ij}]\in \mathbb{R}^{N\times N}$ is defined by $a_{ii}=0$ and $a_{ij}\geq 0$.  Here, $a_{ij}>0$ means there is an edge $(j,\,i)$ in this graph with edge weight $a_{ij}$.   The neighbor set of node $i$ is defined as $\mathcal{N}_i=\{j \mid (j,\, i)\in \mathcal {E}\}$ for $i=1,\,\cdots\,, N$.  A directed path is an alternating sequence $i_{1}e_{1}i_{2}e_{2}{\dots}e_{k-1}i_{k}$ of nodes $i_{l}$ and edges $e_{m}=(i_{m},i_{m+1}) \in\mathcal {E}$ for $l=1,2,{\dots},k$.  If there is a directed path between any two vertices, then the digraph is said to be strongly connected.  The in-degree and out-degree of node $i$ is defined by $d^{\mbox{in}}_i=\sum_{j=1}^N a_{ij}$ and $d^{\tiny\mbox{out}}_i=\sum_{j=1}^N a_{ji}$. The Laplacian of digraph $\mathcal{G}$ is defined as $L\triangleq D^{\mbox{in}}-\mathcal{A}$ with $D^{\mbox{in}}=\mbox{diag}(d^{\mbox{in}}_1,\,\dots,\,d^{\mbox{in}}_N)$.  A digraph is weight-balanced if $d^{\tiny\mbox{in}}_i=d^{\tiny\mbox{out}}_i$ holds for any $i=1,\,\dots,\,N$. 

Note that $L{\bm 1}_N={\bm 0}_N$ for any digraph. If this digraph is weight-balanced, it also holds that ${\bm 1}_N^\top L={\bm 0}_N^\top$ and the matrix  $\mbox{Sym}(L)\triangleq \frac{L+L^\top}{2}$ is positive semidefinite. If this weight-balanced digraph is strongly connected, $0$ is a simple eigenvalue of $\mbox{Sym}(L)$ and all other eigenvalues are positive real numbers. In this case, we order these eigenvalues as $\lambda_1=0<\lambda_2\leq \dots\leq \lambda_{N}$.

\subsection{Convex analysis}\
A function $f\colon \R^m \rightarrow \R $ is said to be convex if for  $0\leq a \leq 1$, 
\begin{align*}
f(a\zeta_1+(1-a)\zeta_2)\leq af(\zeta_1)+(1-a)f(\zeta_2),\, \forall \zeta_1,\zeta_2 \in \R^m
\end{align*}
When the function $f$ is differentiable, it is verified that $f$ is convex if the following inequality holds,
\begin{align*}
f(\zeta_1)-f(\zeta_2)\geq \nabla f(\zeta_2)^\top (\zeta_1 -\zeta_2),\, \forall \zeta_1,\,\zeta_2 \in \mathbb{R}^m 
\end{align*}
and is strictly convex if this inequality is strict whenever $\zeta_1 \neq \zeta_2$.  A function $f$ is $\omega$-strongly convex ($\omega >0$) over $\R^m$ if we have
\begin{align*}
(\nabla f(\zeta_1)-\nabla f(\zeta_2))^\top (\zeta_1 -\zeta_2)\geq \omega |\zeta_1 -\zeta_2|^2,\, \forall \zeta_1,\,\zeta_2 \in \R^m
\end{align*}

A vector-valued function ${\bm f}\colon \R^m \rightarrow \R^m$ is Lipschitz with constant $\vartheta>0$ (or simply $\vartheta$-Lipschitz) if we have
\begin{align*}
|{\bm f}(\zeta_1)-{\bm f}(\zeta_2)|\leq \vartheta |\zeta_1-\zeta_2|,\, \forall \zeta_1, \zeta_2 \in \R^m
\end{align*}

\section{Problem Formulation}\label{sec:formulation}
Consider a multi-agent system consisting of $N$ discrete-time linear agents of the following form:
\begin{align}\label{sys:agent}
\begin{split}
x_{i}(t + 1) &= Ax_{i}(t) + Bu_{i}(t)+d_i(t),~~i= 1,\,2,\,\ldots,\, N\\
y_{i}(t) &= Cx_{i}(t),~~~~t=0,\,1,\,2,\,\ldots
\end{split}
\end{align}
where $x_{i}(t) \in \mathbb{R}^{n_x}$ is the state,  $u_{i}(t) \in \mathbb{R}^{n_u}$ is the input, and $y_{i}(t) \in \mathbb{R}$ is the output of agent $i$. The system matrices $(C,\,A,\,B)$ are assumed to be minimal with compatible dimensions. Here, $d_i(t)\in \R^{n_x}$ represents external disturbances modeled by  
\begin{align}\label{sys:disturbance}
d_i(t)=Ew_i(t),~~~w_i(t+1)=Sw_i(t),~~t=0,\,1,\,2,\,\ldots
\end{align}
where $w_i\in \R^{n_w}$ is the full internal state of external disturbances.   As usual, we assume that $S$ has no eigenvalue inside the unit circle on the complex plane \cite{huang2004nonlinear}. In fact, the components of $w_i$ corresponding to the eigenvalues inside the unit circle will exponentially converge to zero and thus in no way affect the designed goal. 

Each agent is assigned with a local objective function $f_i\colon \R \to \R$. We define an aggregate objective function for this multi-agent system as the sum of these local functions, i.e., $f(s)=\sum_{i=1}^{N} f_i(s)$. This aggregate objective function $f$ is called the global cost function of this multi-agent system.  

Here is an assumption to ensure the existence and uniqueness of minimal solutions to function $f$.
\begin{ass}\label{ass:cost}
	For any $1\leq i\leq N$, the cost function $f_{i}$ is  $\underline{l}$-strongly convex and $\nabla f_{i}$ is $\bar l$-Lipschitz.
\end{ass}

Similar assumptions have been widely used in literature, e.g. \cite{jakovetic2014linear,kia2015distributed,zhang2017distributed,nesterov2018lectures}.  As usual, we assume this unique optimal solution is finite and denote it as $y^{*}$, that is,  
\begin{align}\label{opt:main}
y^{*}=\arg\min_{ s \in \R} \; f(s)\triangleq \sum_{i=1}^{N} f_i(s)
\end{align}

We aim to develop distributed algorithms to drive the outputs of all agents to achieve a consensus on the minimizer $y^{*}$ of $f$ without setting up a centralized working station, which might be expensive or prohibited in some circumstances.  

For this purpose, a weighted digraph $\mathcal{G}=(\mathcal{V},\, \mathcal{E}, \,\mathcal{A})$  is used to describe the information sharing relationships among these agents with a node set $\mathcal{N}=\{1,\,\dots,\, N\}$ and a weight matrix $\mathcal{A}\in \R^{N\times N}$. If agent $i$ can get the information of agent $j$, then there is an edge $(j,\,i)$ in the graph, i.e., $a_{ij}>0$.

The optimal output consensus problem for discrete-time multi-agent system \eqref{sys:agent} can be formulated as follows. {\em Given agent \eqref{sys:agent}, cost function $f_i(\cdot)$, graph $\mathcal{G}$ and disturbance \eqref{sys:disturbance}, the optimal output consensus problem is to find a feedback control $u_i$ for agent $i$ by using its own and neighboring information such that all trajectories of agents are bounded and the resultant outputs satisfy $\lim_{t\to +\infty}|y_i(t)-y^*|=0$ for any  $i=1,\,\dots,\,N$.
}

\begin{rem}\label{rem:formulation}		
	In this formulation, these agents are required to achieve an output consensus minimizing the aggregate global cost function. When agents are all single integrators without disturbances, our formulated problem is coincided with the well-studied distributed optimization or optimal consensus results \cite{nedic2009distributed, lei2016primal}. Here, we further consider multi-agent systems having discrete-time high-order dynamics subject to external disturbances.
\end{rem}

To guarantee that any agent's information can reach any other agents through a directed information flow, we suppose the following assumption is fulfilled as in many publications \cite{kia2015distributed,tang2015distributed,zhang2017distributed}.
\begin{ass}\label{ass:graph}
 $\mathcal{G}$ is strongly connected and weight-balanced.
\end{ass}

Note that when this optimal output consensus problem is solved, we have $\lim_{t\to \infty}y_i(t)=y^*$.  It is natural for agent $i$ to reach some steady state. Thus, we make another assumption to ensure this point.

\begin{ass}\label{ass:re}
	There are constant matrices $X_1,\,X_2$ and $U_1,\,U_2$ with compatible dimensions satisfying that:
	\begin{align}\label{eq:regulator}
	\begin{split}
	X_1&= AX_1 + BU_1,\qquad \quad 1= C X_1\\
	X_2S &= A X_2 + BU_2+E,\quad {\bm 0}^\top= C X_2
	\end{split}
	\end{align}
\end{ass}

Assumption \ref{ass:re} is known as the solvability of regulator equations to achieve set-point regulation and disturbance rejection for discrete-time linear systems \cite{huang2004nonlinear}, which plays a crucial role in resolving our optimal consensus problem. Some well-known verifiable conditions can be found in \cite{huang2004nonlinear}. Under this assumption, we can directly solve the linear matrix equations to obtain their solutions. In this way, one can further obtain the steady-state state and input for each agent as  $X_1y^*+X_1w(t)$ and $U_1y^*+U_2w(t)$  when the optimal output consensus is achieved at $y^*$ with disturbance rejection.

To reject the external disturbances, we suppose the following condition holds without loss of generality.
\begin{ass}\label{ass:observer}
	The pair $\left(  \begin{bmatrix}
	C~~{\bm 0}
	\end{bmatrix},\, \begin{bmatrix}
	A&E\\
	{\bm 0}& S
	\end{bmatrix}\right)$ is observable.
\end{ass}
This assumption implies that the external disturbances can indeed affect our regulated output $y_i$. A sufficient condition to ensure this assumption is the observability of $(E,\, S)$, which can be trivially verified by PBH-test.   

\section{Main Results}\label{sec:main}

To avoid the difficulties brought by the high-order linear dynamics, we develop an embedded design in two steps as that in \cite{tang2018ijrnc, tang2019optimal} to achieve the expected optimal output consensus for these agents over directed graphs. 
 
\subsection{Optimal signal generation}
To begin with, we consider an optimal consensus problem for a group of virtual agents 
\begin{align}\label{sys:abs}
z_{i}(t+1) = z_i(t)+\mu_{i}(t)
\end{align}
with the same cost function $f_i$ and information sharing graph $\mathcal{G}$. 

Under Assumption \ref{ass:cost}, the optimization problem \eqref{opt:main} can be reformulated to the following equivalent form:
\begin{align*}
\min \tilde f(y)\triangleq \sum_{i=1}^N f_i(y_i), ~~ \mbox{subject to } Ly=0
\end{align*}
with $y\triangleq \mbox{col}(y_1,\,\dots,\,y_N)$ and $L$ is the Laplacian of this digraph.   Moreover, the associated Lagrangian of this auxiliary optimization problem is then  $\mathcal{L}(y,\,\tilde \Lambda)=\tilde f(y)+\tilde \Lambda^\top Ly$ with $\tilde \Lambda\in \R^N$.  

There have been some distributed rules to achieve optimal consensus goal or compute the global optimal solution $y^*$ under directed graphs for agent \eqref{sys:abs}, e.g. \cite{kia2015distributed,nedic2017achieving}.  However, most of these algorithms require some initialization process. Since there might be disturbances or round-off errors, such an initialization could fail to be fulfilled during the implementation of these rules. Thus, we are more interested to construct optimal signal generators free of such initializations. 

Note that when the digraph is undirected and connected, the Laplacian $L$ is symmetric and the optimal point can be readily derived by a primal-dual dynamics (e.g. \cite{lei2016primal}).  As for digraphs,  we lose such a symmetry and the original primal-dual method fails to generate the optimal point. Here, we extend the primal-dual dynamic to weight-balanced graphs by adding a proportional terms as follows.
\begin{align}\label{osg}
z_{i}(t+1) &= z_{i}(t) - \gamma(\alpha \nabla f_{i}(z_{i}(t))+\beta Lz_{i}(t) + L\lambda_{i}(t)) \nonumber\\
\lambda_{i}(t+1) &{}= \lambda_{i}(t) + \gamma \alpha \beta Lz_{i}(t)  
\end{align}
where $\alpha,\, \beta$ and $\gamma$ are positive constants to be specified later. This algorithm has been partially investigated in \cite{lei2016primal} when $\alpha=\beta=1$. Here, we add these two tunable parameters to ensure its efficiency with directed graphs. 

Putting system \eqref{osg} into a compact form gives that 
\begin{align}\label{osg-compact}
Z(t+1) &= Z(t)  - \gamma [\alpha \nabla \tilde {f}(Z(t)) + \beta LZ(t)+L\Lambda (t)] \nonumber\\
\Lambda(t+1) &= \Lambda(t) + \gamma \alpha \beta LZ(t)
\end{align}
where $Z(t) = \mbox{col}(z_{1}(t),\,\ldots,\, z_{N}(t))$, $\Lambda(t) = \mbox{col}(\lambda_{1}(t),\, \ldots, \,\lambda_{N}(t))$, and  $$\nabla \tilde{f}(Z(t))\triangleq \mbox{col}(\nabla f_{1}(z_{1}(t),\, \ldots, \,\nabla f_{N}(z_{N}(t)))$$ 
Under Assumption \ref{ass:cost}, the function $\nabla \tilde f$ is $\bar l$-Lipschitz in $Z$.

Let $(Z^*,\, \Lambda^*)$ be an equilibrium point of system \eqref{osg-compact}. The following lemma shows that these agents can achieve the expected optimal consensus at the equilibrium  $(Z^*,\, \Lambda^*)$.
\begin{lem}\label{lem:equilibrium}
	Under Assumptions \ref{ass:cost}--\ref{ass:graph}, we have $Z^*=y^*{\bm 1}_N$.
\end{lem}
\pb At the equilibrium point of system \eqref{osg-compact}, we have
\begin{align*}
 \alpha \nabla \tilde {f}(Z^*)+\beta LZ^*+L\Lambda^*={\bm 0},\quad L Z^*={\bm 0}
\end{align*}

By Assumption \ref{ass:graph}, $LZ^*=0$ implies that there exists a constant $z^*_0$ such that $Z^*=z^*_0 {\bm 1}_N$. Multiplying both sides by ${\bm 1}_N^\top$,  we have ${\bm 1}^\top_N [ \alpha \nabla \tilde {f}(Z^*)+\beta LZ^*+L\Lambda^*]={\bm 1}^\top_N  \nabla \tilde {f}(Z^*)=0$. That is, $\nabla f(z_0^*)=0$. From the strong convexity of cost functions, the optimal solution to problem \eqref{opt:main} is unique. This means that $z^*_0=y^*$ and thus $Z^*=y^*{\bm 1}_N$. The proof is complete.
\pe

To develop effective optimal signal generators, we have to choose  appropriate parameters $\alpha,\,\beta,\, \gamma$ such that the equilibrium of  \eqref{osg-compact} is attractive.   Here is a key lemma to ensure this point. Its proof can be found in \textbf{Appendix}.

\begin{lem}\label{lem:osg}
	Under Assumptions \ref{ass:cost}--\ref{ass:graph}, the trajectory of $z_i(t)$ along system \eqref{osg} exponentially converges to the optimal solution $y^*$ of problem \eqref{opt:main} from any initial value if the chosen parameters satisfy 
	\begin{align}\label{eq:osg-parameter}
	\begin{split}
	&\alpha\geq \max\{1,\,\frac{1}{\underline{l}},\,\frac{2  \bar l^2}{\underline{l}\lambda_2}  \}\\
	&\beta\geq \max\{1,\, \frac{4\alpha^2\lambda_{N}^2}{\lambda_2^2}\}\\
	&0<\gamma<\frac{1}{ \beta^4(\lambda_N^2+\bar l^2)}
	\end{split}
	\end{align}
\end{lem}

\begin{rem}
Condition \eqref{eq:osg-parameter} is only sufficient to guarantee the efficiency of generator \eqref{osg} to reproduce $y^*$, which can be conservative. One may prefer to select these parameters from repeated simulations 
by first increasing $\alpha$, $\beta$ and then decreasing $\gamma$ sequentially.	
\end{rem}

\begin{rem}
	Compared with existing optimal consensus design, the designed generator actually solve a distributed optimization problem under weight-balanced directed graphs. Unlike similar rules in \cite{kia2015distributed,zhang2017distributed,xie2019global}, the developed algorithm is initialization-free and more favorable in large scale networks with varying numbers of agents. 
\end{rem}
\subsection{Solvability of optimal consensus problem}
With the above optimal signal generator, we are going to solve the associated reference tracking and disturbance rejection problem for agent $i$ with reference $z_i(t)$ and disturbance $d_i(t)$. 

Recalling some classical output regulation results \cite{huang2004nonlinear}, a full-information control for each agent to achieve optimal output consensus is written as follows:
\begin{align*}
u_{i}^0(t)=Kx_i(t)+K_1y^*+K_2w_i(t),~~t=0,\,1,\,\dots
\end{align*} 
where $K$ is chosen such that $A+BK$ is Schur stable and $K_1=U_1-KX_1$, $K_2=U_2-KX_2$.  In fact, under this full-information control, we can obtain an error system by letting $\bar x_i(t)=x_i(t)-X_1y^*-X_2w_i(t)$ of the following form:
\begin{align*}
\bar x_i(t+1)=(A+BK)\bar x_i(t),~~~~t=0,\,1,\,\dots
\end{align*} 
along which the regulated output $e_i(t)\triangleq y_i(t)-y^*=C\bar x_i(t)$  converges to zero as $t$ goes to infinity. 

However, the disturbance $w_i(t)$ is not available to us and the global optimal solution $y^*$ is also unknown due to the distributedness of the global cost function. Thus, the above full-information control is not applicable to our problem. 

In Lemma \ref{lem:osg}, we have shown that the  optimal solution $y^*$ can be generated by \eqref{osg} exponentially fast. This motivates us to replace $y^*$ by the reference signal $z_i(t)$. As for the unknown external disturbances,  we can estimate them by observer-based methods to complete the whole design.

To this end, a full-state Luenberger observer is constructed to estimate these disturbances as follows.
\begin{align}\label{sys:observer}
\begin{split}
\tilde x_{i}(t + 1) &= (A+L_1C)\tilde x_{i}(t) + Bu_{i}(t)+E\tilde w_i(t)-L_1y_i\\
\tilde w_i(t+1)&=S\tilde w_i(t)+L_2(C\tilde x_i(t)-y_i) 
\end{split}
\end{align}
where  $L_1$ and $L_2$ are chosen gain matrices with compatible dimensions such that the following matrix is Schur stable.  
$$\tilde A_c\triangleq  \begin{bmatrix}
A+L_1C&E\\
L_2C& S
\end{bmatrix}$$
Note that such $L_1$ and $L_2$ indeed exist under Assumption \ref{ass:observer}. 

Here is a lemma to show the estimation capacity of observer \eqref{sys:observer}.
\begin{lem}\label{lem:observer}
	Under Assumption \ref{ass:observer}, for any $1\leq i\leq N$, the signals $\tilde x_i(t)$ and $\tilde w_i(t)$ along the trajectory of system \eqref{sys:observer} exponentially converge to $x_i(t)$ and $w_i(t)$ as $t$ goes to infinity. 
\end{lem} 
\pb To prove this lemma, we denote $\bar {\tilde x}_i= x_i(t)-\tilde x_i(t)$ and  $\bar {\tilde w}_i= w_i(t)-\tilde w_i(t)$. Jointly using equations \eqref{sys:agent}, \eqref{sys:disturbance}, and \eqref{sys:observer}, we have the following estimation error system for agent $i$:
\begin{align*}
\bar{\tilde x}_i(t+1)&=(A+L_1C)\bar{\tilde x}_i(t)+E\bar{\tilde w}_i(t)\\
\bar{\tilde w}_i(t+1)&=L_2C\bar{\tilde x}_i(t)+S\bar{\tilde w}_i(t) 
\end{align*}
or in a compact form:
\begin{align*}
\begin{bmatrix}
\bar{\tilde x}_i(t+1)\\
\bar{\tilde w}_i(t+1)
\end{bmatrix}=\begin{bmatrix}
A+L_1C&E\\
L_2C& S
\end{bmatrix}\begin{bmatrix}
\bar{\tilde x}_i(t)\\
\bar{\tilde w}_i(t)
\end{bmatrix}
\end{align*}

The above block matrix is exactly $\tilde A_c$, which is Schur stable from the selection of $L_1$ and $L_2$.  Hence, both $\bar {\tilde x}_i= x_i(t)-\tilde x_i(t)$ and  $\bar {\tilde w}_i= w_i(t)-\tilde w_i(t)$ will converge to $\bm 0$ exponentially fast as $t$ goes to infinity. The proof is thus complete. \pe 

Based on the optimal signal generator \eqref{osg} and observer \eqref{sys:observer}, we propose a dynamic controller for agent $i$ as follows. 
\begin{align}\label{ctr:full}
u_{i}(t)&=K\tilde x_i(t)+K_1z_i(t)+K_2\tilde w_i(t) \nonumber\\
\tilde x_{i}(t + 1) &= (A+L_1C)\tilde x_{i}(t) + Bu_{i}(t)+E\tilde w_i(t)-L_1y_i \nonumber\\
\tilde w_i(t+1)&=S\tilde w_i(t)+L_2(C\tilde x_i(t)-y_i)  \nonumber\\
z_{i}(t+1) &= z_{i}(t) - \gamma(\alpha \nabla f_{i}(z_{i}(t))+\beta Lz_{i}(t) + L\lambda_{i}(t))  \nonumber\\
\lambda_{i}(t+1) &= \lambda_{i}(t) + \gamma \alpha \beta Lz_{i}(t),~~t=0,\,1,\,\dots
\end{align}
where matrices $L_1,\,L_2$ and parameters $\alpha,\,\beta,\,\gamma$ are chosen as above.  This control law is distributed in sense of using only its own and neighboring information of each agent.

To show its effectiveness, we derive the new error system under control \eqref{ctr:full}  by some mathematical manipulations as follows:  
\begin{align}\label{sys:error}
\begin{split}
\overline{x}_{i}(t+1) &= (A+BK)\overline{x}_{i}(t)+B\Xi_i(t)\\
e_{i}(t) &= C\overline{x}_{i}(t), ~~~ t=0,\, 1,\, \ldots
\end{split}
\end{align}
where $\Xi_i(t)\triangleq K(\tilde x_i(t)-x_i(t))+K_1(z_i(t)-y^*)+K_2(\tilde w_i(t)-w_i(t))$. It is verified that $\Xi_i(t)$ represents the discrepancy between our actual control effort $u_i(t)$ and its corresponding full-information version $u_i^0(t)$.

It is ready to give our main theorem of this paper.
\begin{thm}\label{thm:main}
	Suppose Assumptions \ref{ass:cost}--\ref{ass:observer} hold. Then the optimal output consensus problem for discrete-time linear multi-agent system \eqref{sys:agent}, \eqref{sys:disturbance}, and \eqref{opt:main} is solved by a distributed controller \eqref{ctr:full}.
\end{thm}
\pb To prove this theorem, we first claim that there exists a constant $c>0$ such that $\limsup_{t\to\infty}|e_{i}(t)|\leq c\limsup_{t\to\infty}|\Xi_{i}(t)|$. Moreover, if $\lim_{t\to\infty} |\Xi_{i}(t)|=0$, we have $\lim_{t\to\infty}e_{i}(t)=0$. This property is a variant of input-to-output stability of system \eqref{sys:error} with input $\Xi_i(t)$ and output $e_i(t)$.

To prove it, we denote $G=A+BK$ for short. By the iteration \eqref{sys:error}, one can obtain that 
\begin{equation*}
\begin{aligned}
\overline{x}_{i}(t) = G ^{t}\overline{x}_{i}(0)+\sum_{j=0}^{t-1}G^{t-1-j}B\Xi_{i}(j),~~ i=1,\,\dots,\,N
\end{aligned}
\end{equation*}
Under Assumption \ref{ass:re}, the regulated output $e_i$ is derived as follows.
\begin{align}\label{eq:output-thm}
\begin{split}
e_i(t)&= CG ^{t}\overline{x}_{i}(0)+ C\sum_{j=0}^{t-1}G^{t-1-j} B\Xi_{i}(j)
\end{split}
\end{align}

Since  $G$ is Schur stable, $\lim_{t\to\infty}CG ^{t}\overline{x}_{i}(0)=0$. To estimate the limit superior of $\{|e_i(t)|\}$, we can neglect this term  without affecting the conclusion. Without loss of generality, we assume $b_i\triangleq \limsup_{t\to\infty}|\Xi_{i}(t)|$ is finite. 

By its definition, for any $\e>0$, there exists a large enough integer $M>0$ such that $||\Xi_{i}(t)|-b_i|\leq \e$ holds for any $t>M$. Splitting the last term of \eqref{eq:output-thm} into two parts gives that
\begin{equation*}
\begin{split}
&|C\sum_{j=0}^{M}G^{t-1-j}B\Xi_{i}(j)+C\sum_{j = M+1}^{t - 1}G^{t-1-j}B\Xi_{i}(j)|\\
&\leq |C||B|[\sum_{j=0}^{M}|G|^{t-1-j}|\Xi_{i}(j)|+\sum_{j=M+1}^{t - 1}|G|^{t-1-j}|\Xi_{i}(j)|]\\
&\leq |C||B||G^{t-1-M}|\sum_{j=0}^{M}|G|^{M-j}|\Xi_{i}(j)|\\
&+(b_i+\e)\sum_{j=M+1}^{t - 1}|G|^{t-1-j}
\end{split}
\end{equation*}
Note that
$$\sum_{j=M+1}^{t - 1}|G|^{t-1-j}  = \frac{1 - |G| ^{t - M}}{1 - |G| } < \frac{1}{1 - |G| }$$ and $|G|^{t-1-M} \to  0$ as $t\rightarrow\infty$ by its Schurness. We have that
\begin{equation*}
\begin{split}
\limsup_{t\to\infty}|e_i|& \leq  \frac{1}{1 - |G| }(b_{i} + \varepsilon)
\end{split}
\end{equation*}
Since  $\varepsilon$ can be chosen arbitrarily, this further implies that $\limsup_{t\to\infty}|e_{i}(t)|\leq c\limsup_{t\to\infty}|\Xi_{i}(t)|$ for $c = \frac{1}{1 - |G| }$. That is, our initial claim is correct.

With the above claim, we only have to show $\Xi_i(t)\to 0$ as $t\to \infty$ under the controller \eqref{ctr:full}, which trivially holds by  Lemmas \ref{lem:osg} and \ref{lem:observer}. Thus, one can conclude that $e_i(t)=y_i(t)-y^*$ converges to $0$ as $t$ goes to infinity. The proof is thus complete.
\pe

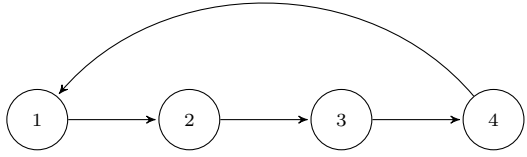
\begin{figure} 
	\centering
	\begin{tikzpicture}[scale=1.5, shorten >=1pt, node distance=2 cm, >=stealth',
	every state/.style ={circle, minimum width=0.8cm, minimum height=0.8cm}]
	\node[align=center,state](node1) {\scriptsize 1};
	\node[align=center,state](node2)[right of=node1]{\scriptsize 2};
	\node[align=center,state](node3)[right of=node2]{\scriptsize 3};
	\node[align=center,state](node4)[right of=node3]{\scriptsize 4};
	\path[->]  (node1) edge (node2)
	(node2) edge (node3)
	(node3) edge (node4)
	(node4) edge [bend right=50]  (node1)
	;
	\end{tikzpicture}
	\caption{Communication graph $\mathcal{G}$ in our example. } \label{fig:graph}
\end{figure}

\begin{rem}
	This theorem can be regarded as a discrete-time companion of existing optimal output consensus results for continuous-time agents derived in \cite{xie2019global,tang2019optimal}. Compared with the well-studied optimal consensus problem for discrete-time single integrators \cite{nedic2009distributed,shi2015extra,lei2016primal,scaman2018optimal}, we extend them to a more general case with high-order linear multi-agent systems subject to nontrivial disturbances. Particularly, we achieve an output average consensus for these linear agents by letting $f_i(s)=(s-y_i(0))^2$ with disturbance rejection.
\end{rem}

\section{Simulations}\label{sec:simu}

In this section, we provide  a numerical example to illustrate the effectiveness of our previous designs.

Consider a multi-agent system including four agents as follows.
\begin{align*}
x_{i}(t + 1) &= \begin{bmatrix}
1 & 1\\
0  & 1 
\end{bmatrix} x_{i}(t) + \begin{bmatrix}
0.5\\
1
\end{bmatrix} u_{i}(t)+d_i(t)\\
y_{i}(t) &=\begin{bmatrix}
1&0
\end{bmatrix} x_{i}(t),~~~~t=0,\,1,\,2,\,\ldots.
\end{align*}
where the external disturbance $d_i(t)$ is generated by an exosystem of the form \eqref{sys:disturbance} with  
\begin{align*}
E=\begin{bmatrix}0.5&0.5\\
\sin(1)-\cos(1)&-\cos(1)-\sin(1)
\end{bmatrix}
\end{align*}
and 
\begin{align*}
S=\begin{bmatrix}
\cos(1)&\sin(1)\\
-\sin(1)&\cos(1)
\end{bmatrix}
\end{align*}
The observability of $(E,\,S)$ can be verified, which implies Assumption \ref{ass:observer}. Assumption \ref{ass:re} is also confirmed with 
\begin{align*}
X_1=\begin{bmatrix}
1\\0
\end{bmatrix},~~U_1=0,~~X_2=\begin{bmatrix}
0&0\\
-1&-1
\end{bmatrix},~~U_2=\begin{bmatrix}
2&2
\end{bmatrix}
\end{align*}

\begin{figure}
	\centering
	\includegraphics[width=0.84\textwidth]{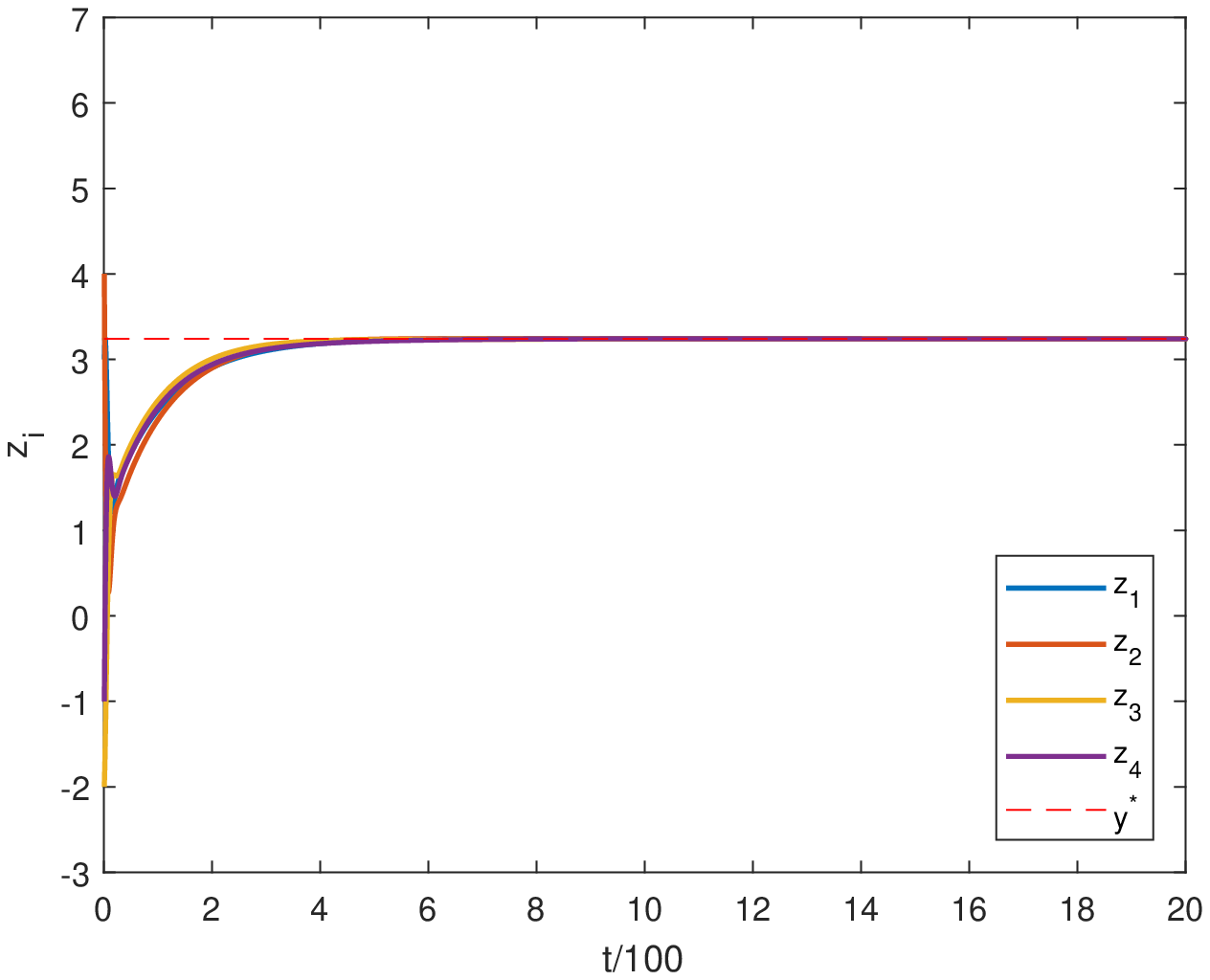}
	\caption{Profiles of $z_i$ under the generator \eqref{osg}.}\label{fig:osg-z}
\end{figure}

\begin{figure}
	\centering
	\includegraphics[width=0.84\textwidth]{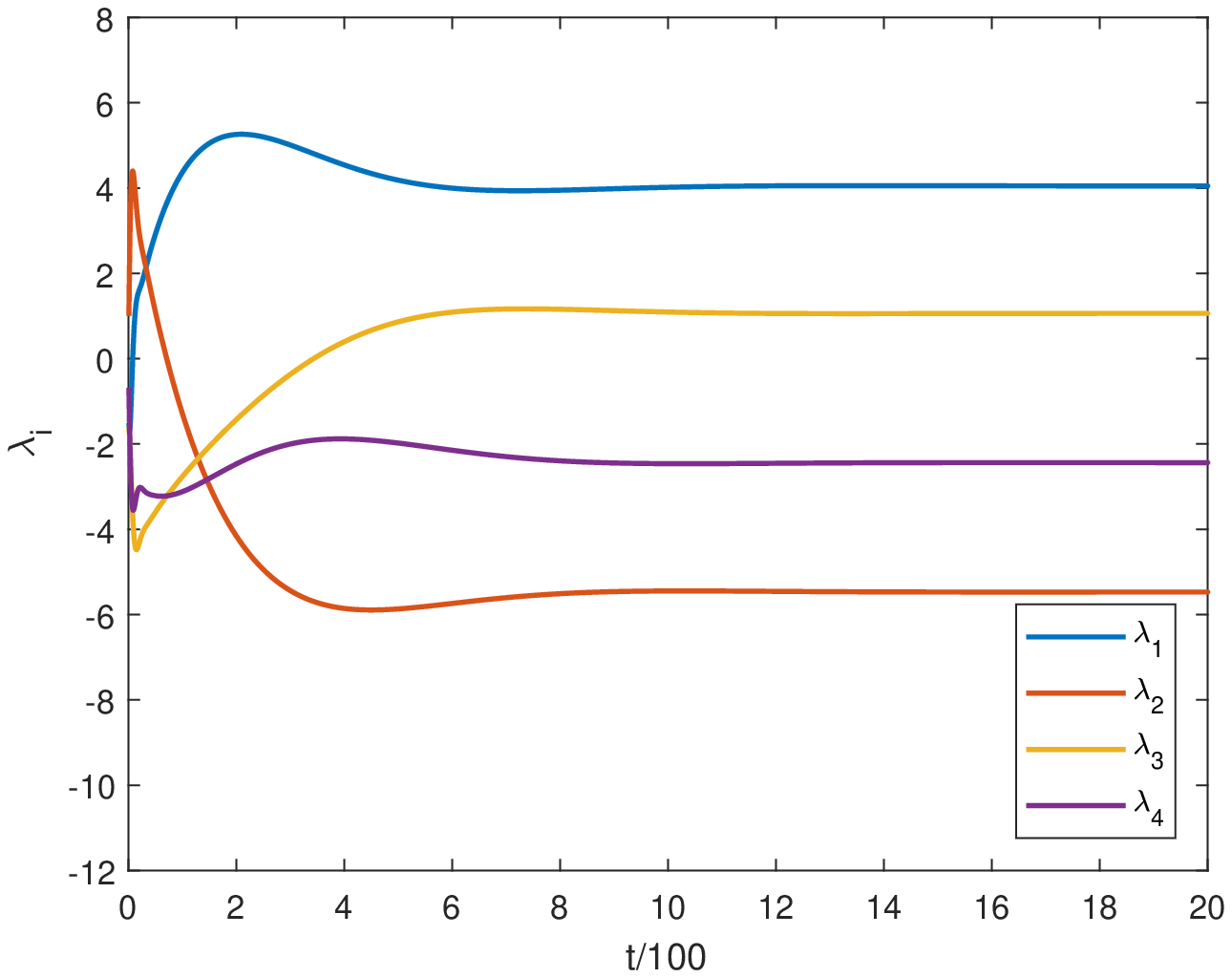}
	\caption{Profiles of $\lm_i$ under the generator \eqref{osg}.}\label{fig:osg-lm}
\end{figure} 
 
The communication topology among these agents is represented by a directed ring graph depicted as Fig.~\ref{fig:graph} with unity edge weights, which satisfies Assumption \ref{ass:graph} with $\lambda_2=1$ and $|L|=2$. The local cost functions are chosen as below.
\begin{align*}
f_{1}(y) &= (y-8)^2\\
f_{2}(y)&= \frac{y^2}{20\sqrt{y^2 + 1}} + y^2\\
f_{3}(y) &= \frac{y^2}{80\ln(y^2 + 2)} + (y-5)^2 \\
f_{4}(y) &= \ln{(e^{-0.005y} + e^{0.005y})} + y^2
\end{align*}
All these functions are strongly convex with Lipschitz gradients. In fact, Assumption \ref{ass:cost} is verified with $\underline{l}=1$ and $\bar l=3$. By minimizing $f(y)=\sum_{i=1}^4f_i(y)$, the global optimal point is $y^{*} = 3.24$.

According to Theorem \ref{thm:main}, the associated optimal output consensus problem can be solved by a control \eqref{ctr:full}. For simulations, we choose  $\alpha=1$, $\beta=15$, $\gamma=0.004$. The state profiles of the developed optimal signal generator are shown in Figs.~\ref{fig:osg-z} and \ref{fig:osg-lm}, where the optimal point $y^*$ can be reproduced quickly while all trajectories of this optimal signal generator keep to be bounded.

\begin{figure*}[!htb]
	\centering
	\includegraphics[width=0.88\textwidth ]{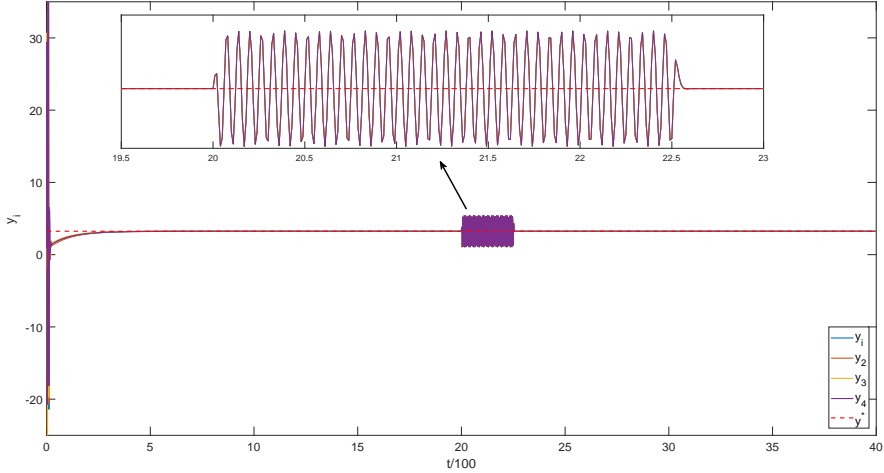}
	\caption{Output trajectories of agents under the controller \eqref{ctr:full}.}\label{fig:oos-c}
\end{figure*}

Next, we choose the gain matrices in control \eqref{ctr:full} as follows.
\begin{align*}
K=\begin{bmatrix}
  -0.4345\\ -1.0285
\end{bmatrix}^\top ,\, L_1=\begin{bmatrix}
 -1.8184\\
-0.3543
\end{bmatrix},\, L_2=\begin{bmatrix}
 -0.1527\\
-0.3141
\end{bmatrix}
\end{align*}

As shown in Fig.~\ref{fig:oos-c}, the optimal output consensus for all agents is achieved on the global optimal solution $y^*$. To make it more interesting, we shut down the disturbance rejection part in the controller (i.e., set $K_2={\bm 0}$) between $t=2000$ and $t=2250$ and find that these agents fail to achieve a consensus. After we restart this part, the optimal output consensus is quickly recovered, which verifies the efficiency of our control to reject these periodic disturbances.

\section{Conclusions}\label{sec:con}
The paper has studied an optimal output consensus for discrete-time linear multi-agent systems subject to external disturbances.  Following an embedded control design, we have employed a primal-dual rule with fixed step sizes to generate the optimal point and developed effective observer-based tracking controllers for these agents to achieve the expected optimal output consensus goal. Future works will consider time-varying digraphs.

\section{Acknowledgments}

This work was supported in part by the National Nature Science Foundation of China under Grant 61973043 and National Key R\&D Program of China under Grant 2018AAA0102600.

\section{Appendix: Proof of Lemma \ref{lem:osg}} \label{app}

First, we denote $r_N=\frac{1}{\sqrt{N}}{\bm  1}_N$, $\Pi_N=I_{N}-r_N r_N^\top$ and let $R_N\in \mathbb{R}^{N \times (N-1)}$ be the matrix satisfying $R_N^\top r_N={\bm  0}_{N-1}$, $R_N^\top R_N=I_{N-1}$ and $R_N R_N^\top=\Pi_{N}$. Note that the matrix $R_N^\top \mbox{Sym}(L) R_N$ is positive definite with eigenvalues $\lambda_2,\,\dots,\,\lambda_{N}$.

We perform  the coordinate transformations: $ \bar Z_1(t)=r_N^\top (Z(t)-Z^*)$, $\bar Z_2(t)=R_N^\top (Z(t)-Z^*)$, $\bar \Lambda_1(t)=r_N^\top \Lambda(t)$, and $\bar \Lambda_2(t)=R_N^\top \Lambda(t)$.  The translated system is given as 
\begin{align*}
\bar Z_1(t+1) &= \bar Z_1(t)-\gamma \alpha  r_N^\top\Delta_1  \nonumber\\
\bar Z_2(t+1) &= \bar Z_2(t)-\gamma R_N^\top [\alpha\Delta_1   +  \beta L\bar Z(t) + L R_N  \bar\Lambda_2 (t)] \nonumber\\
\bar\Lambda_2(t+1) &=\bar \Lambda_2(t) + \gamma \alpha \beta R_N^\top L\bar Z(t)
\end{align*}
where $\bar Z(t)=\mbox{col}(\bar Z_1(t),\,\bar Z_2(t))$ and $\Delta_1 \triangleq \nabla \tilde {f}(Z(t))-\nabla \tilde {f}(Z^*)$. It can be verified that  $\Delta_1 $ is $\bar l$-Lipschitz with respect to $\bar Z(t)$ under Assumption \ref{ass:cost}.

By further letting $\xi(t)=\bar \Lambda_2(t)+ \alpha R_N^\top \bar Z_2(t)$, we have 
\begin{align}\label{osg-compact-trans}
\bar Z_1(t+1) &= \bar Z_1(t)-\gamma \alpha  r_N^\top\Delta_1 \nonumber\\
\bar Z_2(t+1) &= \bar Z_2(t)-\gamma  \alpha R_N^\top \Delta_1 -\gamma  \Delta_2  \nonumber\\
\xi(t+1)&=\xi(t)-\gamma \Delta_3  
\end{align}
where  $R_L=R_N^\top L R_N$, $\Delta_2\triangleq \beta R_L\bar Z_2(t) + R_L \xi(t)- \alpha R_L \bar Z_2(t)$, and $\Delta_3 \triangleq \alpha R_L \xi(t)+\alpha^2 R_N^\top \Delta_1  -\alpha^2 R_L\bar Z_2(t)$.

Next,  we present a Lyapunov analysis to prove the exponential stability of system \eqref{osg-compact-trans}.   Take a quadratic Lyapunov function $V(\bar Z(t),\,\xi(t))=|\bar Z(t)|^2+\frac{1}{\alpha^3}|\xi(t)|^2$ with $\alpha>0$ to be specified later and denote it as $V(t)$ for short.  Apparently, it is positive definite and radially unbounded. The time difference of $V(t)$ along the trajectory of system \eqref{osg-compact-trans} satisfies:
\begin{align*}
\begin{split}
\Delta(t)&\triangleq V(t+1)-V(t)\\
&=|\bar Z_1(t)-\gamma \alpha  r_N^\top\Delta_1 |^2+| \bar Z_2(t)-\gamma  \alpha R_N^\top \Delta_1 -\gamma  \Delta_2 |^2\\
&-|\bar Z(t)|^2+|\xi(t) - \gamma \Delta_3 |^2-|\xi(t)|^2\\
&\leq -2\gamma \alpha \underline{l}|\bar Z(t)|^2+ \gamma^2\alpha^2 |\Delta_1|^2+\gamma^2 |\Delta_2|^2-2\gamma \bar Z_2^\top(t) \Delta_2\\
&-2\gamma^2\alpha \Delta_1^\top R_N \Delta_2- \frac{2 \gamma}{\alpha^3} \xi(t)^\top\Delta_3 + \frac{\gamma^2}{\alpha^3}|\Delta_3|^2
\end{split}
\end{align*}
To handle the above cross terms, we jointly use Young's inequality and the fact $R_N^\top \mbox{Sym}(L)R_N= \frac{R_L+R_L^\top}{2}$. It follows that 
\begin{align*}
-2\gamma \bar Z_2^\top(t) \Delta_2 &=-2\gamma \bar Z^\top_2  [ \beta R_L\bar Z_2(t) + R_L \xi(t)- \alpha R_L \bar Z_2(t)]\\
&\leq -2\gamma\beta\lambda_2 |\bar Z_2(t)|^2+2\gamma\alpha \lambda_{N} |\bar Z_2(t)|^2-2\gamma \bar Z^\top_2 R_L \xi(t)\\
&\leq -\gamma(2\beta \lambda_2-2\alpha\lambda_{N}-\frac{3\alpha^2\lambda_{N}^2}{\lambda_2})|\bar Z_2(t)|^2+\frac{\gamma\lambda_2}{3\alpha^2}|\xi(t)|^2
\end{align*}
and 
\begin{align*}
- 2 \gamma \xi(t)^\top\Delta_3 &\leq -2 \gamma \xi(t)^\top[ \alpha R_L \xi(t)+\alpha^2 R^\top \Delta_1  -\alpha^2 R_L \bar Z_2(t)]\\
&\leq -2\gamma \alpha \lambda_2|\xi(t)|^2-2 \gamma  \alpha^2 \xi(t)^\top R^\top \Delta_1+ 2 \gamma \alpha^2 \xi(t)^\top R_L \bar Z_2(t)\\
&\leq -\frac{4}{3}\gamma \alpha \lambda_2|\xi(t)|^2+ \frac{3\gamma \alpha^3 \bar l^2}{\lambda_2} |\bar Z(t)|^2+ \frac{3\gamma \alpha^3\lambda_{N}^2}{\lambda_2}|\bar Z_2(t)|^2
\end{align*}
Using the above two inequalities, one can derive that
\begin{align*}
\begin{split}
\Delta(t)&\leq -2\gamma \alpha \underline{l}|\bar Z(t)|^2+ \gamma^2\alpha^2 |\Delta_1|^2+\gamma^2 |\Delta_2|^2-2\gamma^2\alpha \Delta_1^\top R \Delta_2\\
&-\gamma(2\beta \lambda_2-2\alpha\lambda_{N}-\frac{3\alpha^2\lambda_{N}^2}{\lambda_2})|\bar Z_2(t)|^2+\frac{\gamma\lambda_2}{3\alpha^2}|\xi(t)|^2\\
&-\frac{4\gamma \lambda_2}{3\alpha^2}|\xi(t)|^2+ \frac{3\gamma \bar l^2}{\lambda_2} |\bar Z(t)|^2+ \frac{3\gamma \lambda_{N}^2}{\lambda_2}|\bar Z_2(t)|^2 +\frac{\gamma^2}{\alpha^3}|\Delta_3|^2\\
&\leq -\gamma(2\alpha \underline{l}-\frac{3  \bar l^2}{\lambda_2} )|\bar Z(t)|^2-\frac{\gamma \lambda_2}{\alpha^2}|\xi(t)|^2\\
&-\gamma(2\beta \lambda_2-2\alpha\lambda_{N}-\frac{3\alpha^2\lambda_{N}^2}{\lambda_2}-\frac{3\lambda_{N}^2}{\lambda_2})|\bar Z_2(t)|^2\\
&+2\gamma^2\alpha^2 |\Delta_1|^2+2\gamma^2 |\Delta_2|^2+\frac{\gamma^2}{\alpha^3}|\Delta_3|^2
\end{split}
\end{align*}

Letting $\alpha\geq \max\{1,\,\frac{1}{\underline{l}},\,\frac{2  \bar l^2}{\underline{l}\lambda_2}  \}$ and $\beta\geq \max\{1,\, \frac{4\alpha^2\lambda_{N}^2}{\lambda_2^2} \}$ gives 
\begin{align*}
\Delta(t)&\leq -\frac{\gamma}{2}V(t)+2\gamma^2\alpha^2 |\Delta_1|^2+2\gamma^2 |\Delta_2|^2+\frac{\gamma^2}{\alpha^3}|\Delta_3|^2
\end{align*}

We use Young's inequality again to dominate the last three terms and obtain the following relationships: 
\begin{align*}
|\Delta_1|^2&\leq \bar l^2 |\bar Z(t)|^2\\
|\Delta_2|^2&\leq 2\max\{(\beta-\alpha)^2, 1\} \lambda_N^2(|\bar Z_2(t)|^2+|\xi(t)|^2)\\
&\leq 2\beta^2\lambda_N^2(|\bar Z_2(t)|^2+|\xi(t)|^2)\\
\frac{|\Delta_3|^2}{\alpha^3}&\leq \frac{1}{\alpha}|R_L \xi(t)+\alpha R^\top \Delta_1  -\alpha R_L\bar Z_2(t)|^2\\
&\leq \frac{3\lambda_{N}^2}{\alpha}|\xi(t)|^2+3\alpha\bar l^2 |\bar Z(t)|^2+ 3\alpha   \lambda_{N}^2|\bar Z_2(t)|^2
\end{align*}
Combining these inequalities gives
\begin{align*}
\Delta(t)&\leq -\frac{\gamma}{2}V(t)+2\gamma^2\alpha^2 \bar l^2 |\bar Z(t)|^2+ 4\gamma^2\beta^2\lambda_N^2|\bar Z_2(t)|^2+4\gamma^2\beta^2\lambda_N^2|\xi(t)|^2+\frac{3\gamma^2\lambda_{N}^2}{\alpha}|\xi(t)|^2\\
&+3\gamma^2\alpha\bar l^2 |\bar Z(t)|^2+ 3\gamma^2\alpha \lambda_{N}^2|\bar Z_2(t)|^2\\ 
&\leq -\frac{\gamma}{2}V(t)+5\gamma^2\beta^2(\lambda_N^2+\bar l^2)(|\bar Z(t)|^2+|\xi(t)|^2)\\
&\leq -\frac{\gamma}{2}V(t)+5\gamma^2\beta^2(\lambda_N^2+\bar l^2)\alpha^3V(t)\\
&\leq -\frac{\gamma}{2}V(t)+ \frac{5}{16}\gamma^2\beta^4(\lambda_N^2+\bar l^2)V(t)
\end{align*}
By setting $0<\gamma<\frac{1}{ \beta^4(\lambda_N^2+\bar l^2)}$, it follows that 
\begin{align*}
\Delta(t)&\leq -\frac{3\gamma}{16}V(t)
\end{align*}
According to Theorem 2 in \cite{jiang2002converse},  we obtain the exponential convergence of $V(t)$ and thus $\bar Z(t)$ to the origin as $t\to \infty$. By Lemma \ref{lem:equilibrium}, the proof is complete.

\end{document}